\documentclass[aps,floats,twocolumn,epsf,prb,showpacs]{revtex4-1}
\usepackage{graphicx} % Include figure files
\usepackage{graphics}
\usepackage{amsmath}
\begin{document}

\let\n=\nu
\let\o=\omega
\let\s=\sigma
\def\np{\n'}
\def\sp{\s'}
\def\EL{E_{L}}
\def\EN{E_N}
\def\ES{E_S}
\def\EM{E_M}
\def\ExN{\mbox{e}^{-\beta \EN}}
\def\ExM{\mbox{e}^{-\beta \EM}}
\def\ExL{\mbox{e}^{-\beta \EL}}
\def\ExS{\mbox{e}^{-\beta \ES}}
\def\c{(c_{\s})}
\def\cd{(c_{\s}^{\dagger})}
\def\cp{(c_{\sp})}
\def\cpd{(c_{\sp}^{\dagger})}

\title{
 Electronic structure of nickelates: from 2D heterostructures to 3D bulk materials
}\author{P. Hansmann$^1$,  A. Toschi$^1$, Xiaoping Yang$^2$,
O. K. Andersen$^2$, K. Held$^1$}

\affiliation{$^1$ Institute for Solid State Physics, Vienna University of Technology, 1040 Vienna, Austria\\
$^2$ Max-Planck-Institut f\"ur Festk\"orperforschung, 70569}

\date{\today}

\begin{abstract}
Reduced dimensionality and strong electronic correlations, which are
among the most important ingredients for cuprate-like high-T$_{\rm c}$
superconductivity, characterize also the physics of nickelate based
heterostructures. Starting from the local density approximation we arrive
at a simple two-band model for quasi two-dimensional (2D)
LaNiO$_3$/LaAlO$_3$ heterostructures and extend it by introducing an
appropriate hopping in the $z$ direction to describe the dimensional
crossover to three dimensions (3D). Using  dynamical mean field
theory, we study the effects of electronic correlations with
increasing interaction strength along the crossover from 2D to
3D. Qualitatively, the effects of electronic correlations are
surprisingly similar, albeit quantitatively larger interaction
strengths are required in three dimensions for getting a Mott-Hubbard
insulating state. The exchange parameters of an effective
Kugel-Khomskii-like spin--orbital--model are also derived and reveal
strong antiferromagnetic tendencies. 
\end{abstract}

% 71.27.+a  Strongly correlated electron systems; heavy fermions
% 71.10.Fd  Lattice fermion models (Hubbard model, etc.)
% 71.30.+h  Metal-insulator transitions and other electronic transitions
% 71.20.Eh  Rare earth metals and alloys
% 75.20.Hr  Local moment in compounds and alloys; Kondo effect, valence
%           fluctuations, heavy fermions
% 75.47.Gk  Colossal magnetoresistance

\maketitle
\section{Introduction}
With the tremendous experimental progress \cite{hetero} in growing oxide heterostructures, new routes to engineer the dimensionality of materials have become available. A physical phenomenon particularly sensible to dimensionality is high--T$_{\rm c}$ superconductivity \cite{Dagotto94} which seems to find its ideal playground in quasi two--dimensional systems. Maybe not accidentally, in two dimensions also strong antiferromagnetic spin fluctuations prosper, without undergoing a phase transition to long-range order as in three dimensions.

Besides two dimensionality, the electronic structure of the most prominent class of unconventional superconductors, the cuprate family, is governed by a single Cu $d_{x^2-y^2}$ orbital doped slightly away from half--filling. 
An important step towards a better understanding of cuprate superconductivity was taken by Pavarini \emph{et al.} \cite{Pavarini} who found an empirical trend in the material dependence of the critical temperature, T$_{\rm c}$, at optimal doping. In the proposed scenario the material dependent parameter is the energy difference between the \emph{planar} Cu $d_{x^2-y^2}$ orbital at the Fermi energy and an \emph{axial} Cu $4s$-like orbital at higher energy. 
%Also important, in this respect, seems to be the ratio ***$r=t'/t$ (this is true only for small r)*** 
%of next-nearest neighbor (n.n.n.) hopping $t'$ to nearest neighbor (n.n.) hopping $t$ in the
%single orbital \cite{Pavarini}. At least the cuprates with the highest critical temperature 
%$T_c$ have $r\sim 0.4$.  This ratio in turn can be understood from the interplay, or the energy distance, between the planar $d_{x^2-y^2}$ and the
%axial Cu $4s$ orbital.

In a recent Letter \cite{chaloupka08}, Chaloupka and Khaliullin proposed the artificial engineering of nickelate heterostructures with the aim of getting cuprate-like electronic structure and interplay between axial and planar orbitals. A possible realization are LaNiO$_3$=LaO-NiO$_2$ layers alternated with insulating layers such as LaAlO$_3$=LaO-AlO$_2$ with the same sequence of sub-layer charges. To avoid macroscopic polarization, the substrate on which the heterostructure is grown should preferably have the same sequence of sub-layer charges (SrTiO$_3$ may for instance not be suitable).
We have shown that in these d$^7$ heterostructures, the Ni $d_{3z^2-r^2}$ orbital can play the role of the \emph{axial} orbital\cite{Hansmann09}; it is pushed considerably above the planar Ni $d_{x^2-y^2}$ orbital as an effect of either the Coulomb interaction, an axially compressive strain, or control of the apical cation (substitution of Al by, e.g., Sc)\cite{Yang10} so that a single Fermi surface sheet with similar geometry and volume as in cuprates emerges, before the system becomes eventually insulating.

The main focus of this complementary paper is a detailed analysis of the Mott-Hubbard transition taking place in the \emph{axial}/\emph{planar} two--band system and of the changes observed when going from the two dimensional heterostructure to three dimensional bulk nickelates. The latter can be viewed as a dimensional crossover, in which the Cu $d_{3z^2-r^2}$ degrees of freedom are more equivalent to the Cu $d_{x^2-y^2}$ ones due to the lack of dimensional constraint. 

For the purpose of this study, we simplify the local-density-approximation (LDA) bandstructure of Ni $e_g$ orbitals by including merely the hopping between the first- and second-nearest neighbor Ni d(e$_g$) Wannier orbitals, in contrast to Ref.~\cite{Hansmann09} where the full LDA bandstructure was considered. This gives a quarter-filled two-band model for the $e_g$ degrees of freedom, i.e., one electron in the two Ni $e_g$ orbitals. We recall here that in transition metal oxides the transition metal ion is usually surrounded by an oxygen octahedron and, depending on the specific case, either $t_{2g}$ (as, e.g., in vanadiumsesquioxide) states or $e_g$ states (as in the nickelates and cuprates) play the most important role in determining the low energy excitations of the system. Hence, our study of an $e_g$--band model at quarter filling and the dimensional crossover from quasi--two-- to three-dimensional systems is of interest also from a more general perspective. 
Moreover, since there are plenty of parameters governing the physics such as the bandwidth of the two bands, their crystal field splitting, the interband hybridization, the Coulomb interaction, and last but not least the dimensionality, it is of great importance to derive such models from realistic bandstructures.

This way we also make  some contact to previous  multi-band model
studies
\cite{heldarita,moeller95,bulla99,rozenberg99,koga05,ferrero05,medici05,poteryaev07}
where features such as the orbital selective Mott transitions
\cite{anisimov02,bluemer09,medici09} were analyzed. Many of these multi--band
studies made approximations, such as neglecting the hybridization
between the orbitals. This poses the question in how far the employed
models are applicable to real world materials. Moreover, most of the
aforementioned studies concentrated on half filling.

In Section \ref{Sec:model}, we will briefly discuss the two-band model and how we derive its parameters. In Section \ref{Sec:MIT}, we discuss the evolution of the spectral function, self energy and double occupation with increasing Coulomb interaction strength. In Section \ref{Sec:3d}, we turn to the dimensional crossover, comparing the quasi two-dimensional system for a heterostructure and a more homogeneous 3D model. Finally, we discuss the issue of magnetic order at low temperatures and derive an effective spin-orbital superexchange model in Section \ref{Sec:spinfluct}, before we arrive at the conclusion in Section \ref{Sec:conclusion}. 

\maketitle

\section{LDA+DMFT }
\label{Sec:model}

\subsection{Effective two-band model}
Density functional calculations  were performed for the simplest, 1/1 unstrained superlattice
LaNiO$_{3}$/LaAlO$_{3}\,$=$\,$LaO-NiO$_{2}$-LaO-AlO$_{2}$ of which a
detailed discussion can be found in Ref.~\onlinecite{Hansmann09}. Here we
recall the most important facts relevant for the discussion
of our model study. In the LaNiO$_{3}$/LaAlO$_{3}$ heterostructure
nickel is in a threevalent state Ni$^{3+}$. The coordination polyhedron
is an octahedron with a small tetragonal distortion due to the
LaAlO$_{3}$ layers along the c-axis. The cubic crystal field eigenstates of the
Nickel d--electrons are the lower lying t$_{2g}$ triplet and the
higher lying $e_g$ doublet. The cubic splitting is large enough so that
the nominal filling in Ni$^{3+}$ is $t_{2g}^6$ (completely filled) and
$e_g^1$ (quarter filling). As a consequence we find two $e_g$ bands at
quarter filling around the Fermi energy to be the relevant degrees of
freedom for the low energy excitations of the system
\cite{Hansmann09}. The insulating LaAlO$_{3}$ layers give rise to the
tetragonal distortion which splits the axial $|3z^2-r^2\rangle$ orbital $\approx 150$meV above the
$|x^2-y^2\rangle$ orbital and, more
importantly, strongly reduce the width of the axial
$|3z^2-r^2\rangle$ band by blocking the perpendicular hopping\cite{Hansmann09}.

In order to obtain the model for our DMFT study we construct a basis of localized Ni $e_g$ Wannier functions using the Nth-order muffin-tin orbital method (NMTO) like in Ref.~\onlinecite{Hansmann09}, but now truncate the hoppings after those between second-nearest Ni neighbors. 
For our purposes of model calculations we prefer this simpler
picture of the systems kinetics in real space and pay the price of
a slightly less accurate representation of the dispersion of the LDA conduction bands. Let us remark that this simplification could affect the Fermi surface transition discussed in Ref.~\onlinecite{Hansmann09} but not the overall properties of the MIT and the 2D to 3D crossover.

In the basis $\{ |x^2-y^2\rangle , |3z^2-r^2\rangle\}$ of the planar and axial $e_g$ Bloch-summed orbitals, the one-electron part of the Hamiltonian is:\\

\begin{multline}\label{MOD_disp2D}
	  \varepsilon^{2D}_\mathbf{k}=-2(\cos k_x+\cos k_y)\cdot\begin{pmatrix}
	  0.45 & 0 \\
	  0    & 0.17
	  \end{pmatrix}\\[0.2cm]
	  +2(\cos k_x-\cos k_y)\cdot\begin{pmatrix}
	  0    & 0.28 \\
	  0.28 & 0\\
	  \end{pmatrix}\\[0.2cm]
          -4(\cos k_x\cdot\cos k_y)\cdot\begin{pmatrix}
	  0.09 & 0 \\
	  0    & 0.03
	  \end{pmatrix}
	  +\begin{pmatrix}
	    0    & 0 \\
	    0    & 0.15
	  \end{pmatrix}
\end{multline}\\

Here, energies are in eV. The first and second terms account for respectively the intra- and inter-orbital nearest-neighbor hoppings. We see that the hopping between the $x^2-y^2$ orbitals is almost three times as large as the one between the $3z^2-r^2$ orbitals and that the inter-orbital hopping  is nearly the geometrical average of the intra-orbital hoppings. This, as well as the factor 3, is what one expects when the Ni-Ni hopping is exclusively via O, i.e., when $t_{\text{Ni-Ni}} = t_{\text{Ni-O}}\cdot t_{\text{O-Ni}}/(\varepsilon_{\text{Ni}}-\varepsilon_{\text{O}})$ and  $t_{\text{Ni-O}}$ follows the canonical rules~\cite{Andersen78}, which in the present case yield: $t_{x^2-y^2} = \sqrt{3}\,t_{3z^2-r^2}$ in $x$-, and $y$-direction. The third term in equation \eqref{MOD_disp2D} is the second-nearest neighbor hopping, and the fourth is the on-site term, i.e. it gives the tetragonal crystal-field splitting. The two last-mentioned terms do not mix the two orbitals.  
In real space, the Hamiltonian of our effective model reads:
\begin{equation}\label{eq:Heff}
  \hat{H}_{\text{mod.}}=\sum_{iljm,\sigma}t_{iljm}c^\dag_{il\sigma}c_{jm\sigma}+\hat{H}_{\text{int.}},
\end{equation}
where $l,m\in\{p,a\}$ label the planar and axial orbitals.
The hopping amplitudes $t_{iljm}$ are the matrices in
Eq.~\eqref{MOD_disp2D} with $l,m$ as the orbital index and $i,j$
as site index of nearest or next nearest neighbors on the square lattice.

\begin{figure}
\begin{center}
  \includegraphics[width=7.5cm]{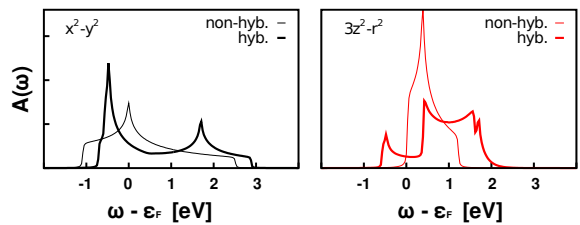}
  \caption{(Color online) Orbitally projected DOS (spectral density) for each of the two $e_g$ bands (thick lines). Neglecting the inter--orbital hybridization leads to the results shown as thin lines.
The hybridization is seen to strongly enhance the width of the $3z^2-r^2$ spectral density, leading to a remarkable itinerancy,
despite the quasi two-dimensional nature of the
LaNiO$_3$/LaAlO$_3$ heterostructure.}
  \label{MOD_nonintDOS}
\end{center}
\end{figure}

The effects of the inter-orbital hybridization are clearly visible in Fig.~\ref{MOD_nonintDOS} showing the
noninteracting density 
of states resulting from the noninteracting part \eqref{MOD_disp2D} of Hamiltonian (\ref{eq:Heff}). A comparison to the case without 
hybridization (t$_{pa}=0$) plotted as a thin line shows a big difference especially for the
$3z^2-r^2$-projection. Its width is strongly enhanced if the hopping to
the more mobile $x^2-y^2$-orbitals is properly taken into
account. This effect is important since otherwise the  reduced hopping
in the $z$-direction would more severely reduce the itinerancy
of electrons in the  $3z^2-r^2$ orbital\cite{footnote01}.
That is, in the absence of the $x^2-y^2$-to-$3z^2-r^2$ hybridization the
$3z^2-r^2$ orbitals would localize too easily (even at a small Coulomb interaction strength).
This scenario was originally proposed in Ref.~\cite{Chaolupka09} prior to realistic material calculations.

The Coulomb interaction
terms read:
\begin{equation}\label{eq:Hint}
\hat{H}_{\text{int.}}=U\sum_{il\sigma}n_{il}^{\uparrow}n_{il}^{\downarrow}+\sum_{il>
  m\sigma\sigma'}(V-\delta_{\sigma,\sigma'}J)n_{il}^{\sigma}n_{im}^{\sigma'}
\end{equation}
where $n_{il}^{\sigma}=c^\dag_{il\sigma}c_{il\sigma}$,
$\delta_{\sigma,\sigma'}$ denotes the Kronecker symbol, U and V represent the
intra-orbital and inter-orbital repulsion, respectively, and J is the Hund
exchange term.

For the solution of the impurity problem in the DMFT self consistency loop we
use quantum Monte Carlo (QMC) simulations of the Hirsch and Fye type\cite{hirschfye} where neither the pair hopping term
(corresponding to highly excited states) nor the spin-flip term (which poses
a severe sign problem in the numerical simulation\cite{Held98a}) have been included.
Presented results were obtained at $\beta=10$eV$^{-1}$ except for the double occupancies shown in the right panel of Fig.~\ref{MOD_NdPlot}. 
Further, convergence with respect to the value of the Trotter discretization $\Delta\tau$ was validated. Afterwards, the maximum entropy method was employed to perform analytic continuation of the Green function to the real axis.
Concerning the choice of the interaction parameter we
perform a systematic analysis where U is varied in a regime of
physically meaningful values in order to investigate the
Mott--transition and its specific features. The Hund's exchange
term, on the other hand, has been fixed to $J=0.7$eV, a reasonable
value close to that of atomic Ni. Moreover, assuming the deviation from cubic
symmetry to be negligible with respect to the Coulomb interaction parameters, we
employ $V=U-2J$.

Before presenting our numerical results a comment about the applicability of DMFT to the finite dimensional system considered is due.
While it has been shown that corrections arising from spatial correlations beyond DMFT \cite{toschi07,toschi09} are generally small in 3D (except in the close proximity of a second order phase transition), stronger effects have to be expected for 2D systems. The strength of such correction with respect to DMFT is, however, decreasing with increasing temperatures \cite{toschi09,Hushcroft01} and therefore can be neglected as a first approximation in the temperature regime we are considering here. This corresponds to the assumption that we are well above any ordering temperature at $\beta=10-25$eV$^{-1}$. Naturally a different treatment to include effects of spatial correlations is necessary in order to study the low temperature regime where spin fluctuations and, possibly, superconductivity emerge. For this reason we include in Sec.~\ref{Sec:spinfluct} a two-site calculation to estimate the size of non-local antiferromagnetic correlations for the system considered.

\subsection{Mott transition in two dimensions}
\label{Sec:MIT}

\subsubsection*{Spectral Function and Self Energy}
\begin{figure}
  \begin{center}
  \includegraphics[width=7.5cm]{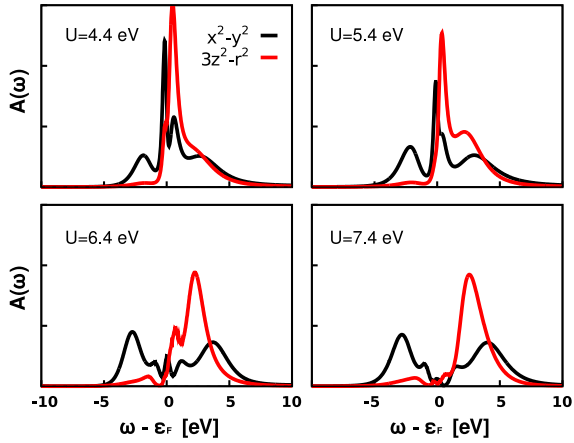}
  \caption{(Color online) Evolution of the LDA+DMFT 
  spectrum with increasing Coulomb interaction $U$ (with $J=0.7$eV, $V=U-2J$) at $\beta=10$eV$^{-1}$.
  The black curves show the $x^2-y^2$ and the red/gray ones the $3z^2-r^2$-projected spectral density.}
  \label{MOD_spectra2D}
  \end{center}
\end{figure}

Fig.~\ref{MOD_spectra2D} shows the evolution of the $k$--integrated
spectral functions $A(\omega)$ for both orbitals with increasing $U$. Let us first focus on  the $x^2-y^2$-projected spectral density (black curve). Already at  $U=4.4$ eV ($V=3.0$ eV; first
panel), the spectrum is strongly renormalized  with respect to the noninteracting
case: A narrow quasiparticle peak at $\varepsilon_{\rm F}$ and Hubbard bands which are separated by an energy of the order
of U are well visible. As the value of the interaction parameter
is increased to $U=5.4$ eV and $U=6.4$ eV (second and third panels),
such spectral features become even more pronounced as one observes a
stronger renormalization of the quasiparticle peak and more pronounced
Hubbard bands. However, at $U=7.4$
eV (fourth panel) a qualitative change has occurred: the 
spectral weight at the Fermi level has vanished and a Mott-Hubbard gap
has been formed. Hence, a Mott metal--insulator
transition has taken place between $U=6.4$ and $7.4$ eV.

Fig.~\ref{MOD_spectra2D} also indicates that the metal--insulator transition of the  $3z^2-r^2$-band (red curve) occurs for
the same critical interaction of $U\gtrsim 6.4$ eV as in the $x^2-y^2$ band. It is, however, qualitatively
different. For $U$ increasing from $4.4$ eV, one observes that the
$3z^2-r^2$ quasiparticle peak is shifted above that of the $x^2-y^2$ peak,
which is at the Fermi level. This is the Coulomb enhancement of the
tetragonal crystal field which causes the Fermi surface to have only
one, mostly $x^2-y^2$--like sheet as in the cuprates\cite{Hansmann09}. For increasing U,
the $3z^2-r^2$ spectral weight shifts further and further above the Fermi
level where for $U=7.4$ eV it has completely disappeared.

The
remaining spectral weight below the Fermi energy, corresponding to the
lower Hubbard band, is entirely due to the hybridization with
the lower $x^2-y^2$-like Hubbard band.

\begin{figure}
\begin{center}
  \includegraphics[width=7.5cm]{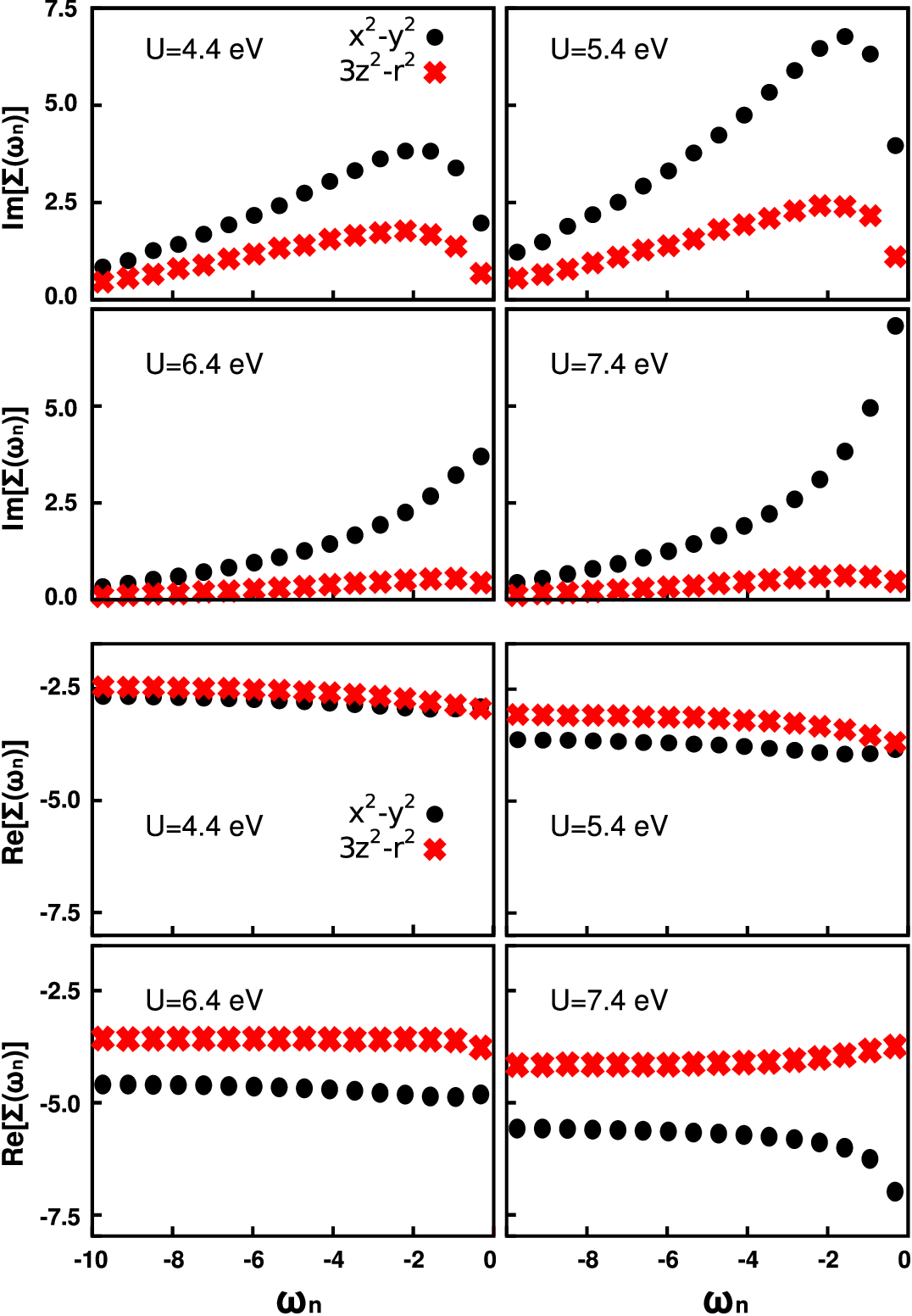}
  \caption{(Color online) Evolution of the LDA+DMFT 
 self energy (vs.\ negative Matsubara frequencies) with increasing $U$ at $\beta=10$eV$^{-1}$.
  The black curves show the $x^2-y^2$ and the red/gray ones the $3z^2-r^2$
  self energy. Via extrapolation of the imaginary and real part of the
  self energy to $\omega_{\rm n}\rightarrow 0$ we can obtain the
  quasiparticle renormalization Z and the ``crystal--field'' enhancement,
  respectively, shown in Fig.~\ref{MOD_muZ2D}.} 
  \label{MOD_self2D}
  \end{center}
\end{figure}

\begin{figure}
\begin{center}
  \includegraphics[width=7.5cm]{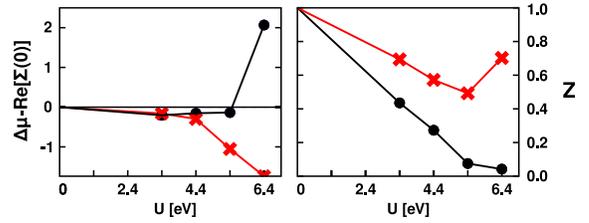}
  \caption{(Color online) Left hand side: Effective chemical potential $\mu-{\rm Re}\Sigma(0)$
    relative to the noninteracting chemical potential for the
    $x^2-y^2$ (black) and $3z^2-r^2$ (red/gray) orbital at $\beta=10$eV$^{-1}$. At the Mott
    transition, the low frequency effective chemical potential of the
    $3z^2-r^2$ orbital is strongly reduced so that it is shifted
    above the Fermi energy, i.e., the $3z^2-r^2$ orbital becomes ``band''--insulating. Right hand side: Quasiparticle weight $Z$  for the
    $x^2-y^2$ (black) and $3z^2-r^2$ (red) orbital (also at $\beta=10$eV$^{-1}$). For the $x^2-y^2$
    orbital, we find  $Z\rightarrow 0$, i.e., the (central) quasiparticle peak
    disappears and the $x^2-y^2$ orbital becomes Mott--insulating.}
  \label{MOD_muZ2D}
\end{center}
\end{figure}

The different nature of the metal--insulator transition in the $x^2-y^2$- and the
$3z^2-r^2$--bands can be identified even more clearly from an analysis of the
self energy. In Fig.~\ref{MOD_self2D}, we show the evolution of the
imaginary part of the self energy
$\text{Im}\Sigma_{x^2-y^2}(\omega_n)$ with increasing
interaction. Again, for the $x^2-y^2$ band, the expected behavior of a
standard Mott transition is found: The appearance of the insulating
gap results from  the divergence of
$\text{Im}\Sigma_{x^2-y^2}(\omega)$ in the limit of zero frequency. In this limit the quasiparticle renormalization factor
$Z=(1-\frac{\partial \text{Im}\Sigma(i \omega)}{\partial\omega}|_{\omega=0})^{-1}$ goes
to zero as a further hallmark of the Mott character of the transition in the
$x^2-y^2$-band (see black circles in the left panel of
Fig.\ref{MOD_muZ2D}). At the same time, however, 
the insulating transition of the $3z^2-r^2$--band is not associated to any qualitative change in
$\text{Im}\Sigma_{3z^2-r^2}(\omega)$. For all interactions  a ``metallic''--like bending
can be observed in $\text{Im}\Sigma_{3z^2-r^2}(\omega)$ at low frequencies. As a
consequence of this, the renormalization factor $Z$ stays always finite (red
crosses in the right panel of Fig.\ref{MOD_muZ2D}). The
disappearance of spectral weight at the Fermi energy in this case is
determined by a large relative shift of the orbitals induced by
the Coulomb interaction. This shift $\Delta_{CF}^{\text{eff}}$, is a consequence
of the strong variation of the real part of the self energies at zero
frequency shown in the left panel of Fig.~\ref{MOD_muZ2D}. More
specifically, while the effective chemical potential of the
$x^2-y^2$-band (defined as $\mu -{\rm Re} \Sigma_{x^2-y^2}(0)$) remains
close to the noninteracting value up to $U=6.4\,$eV,
its value for the $3z^2-r^2$-band is clearly decreasing with
increasing $U$. As a result, the metal insulator transition of the
$3z^2-r^2$-band occurs for a value of $U\gtrsim 6.4$ eV, when the
low energy spectral weight of this band is shifted above the Fermi energy. 

A key feature in the physics we have observed here is the interplay between the
two orbitals stemming from the fairly large hybridization in
equation~\eqref{MOD_disp2D} proportional to $\cos k_x -\cos k_y$. An evident sign of such interplay is the concomitance of the
metal insulator transition in both orbitals. For understanding the
underlying physics it is instructive to consider the different
possible scenarios for the MIT in a two band system with hybridization.

Increasing the Coulomb interaction has basically two effects (see also
Refs.~\onlinecite{keller04}, \onlinecite{paper2} and in particular Ref.~\onlinecite{poteryaev07}): Double occupancy of a site becomes more expensive
and on-site energies of the two orbitals are shifted with respect to each
other. This gives rise to three possible transition scenarios. (i) If
the shift between the bands is increasing slowly with interaction,
double occupancy becomes practically forbidden for a value of V where
both bands are still partially filled. As a result, a
simultaneous Mott transition occurs with $Z\rightarrow 0$ for
{\em both} bands. If, instead, the shift is increasing more
rapidly with interaction, at a certain point one of the
bands will empty, leaving the other one half filled. The
nature of the transition then depends on the value of U for which this happens. 
There are two
possibilities: (ii) If U is strong enough to forbid
double occupancy in the half-filled band, this band will become
insulating  abruptly. Such a situation is reflected in simultaneous, but qualitatively different, 
metal--insulator transitions for both bands. This is what
we observe in Figs. \ref{MOD_spectra2D} and \ref{MOD_self2D}. (iii) The other possibility is
that U is not yet strong enough to prevent the double
occupancies in the half-filled band. The Mott transition in this band
-- differently from our case -- would then not take place simultaneously
with the depletion of the other band. This case is different from the one considered
in Figs. \ref{MOD_spectra2D} and \ref{MOD_self2D} and the only one
in which the metal--insulator--transition can be fully described in terms of
an effective single-band model.

The initial lifting of the orbital degeneracy (the LDA crystal-field splitting)
is expected to push the system in the
direction of situation (ii), which we observed for our model with an
initial crystal field splitting of
$\varepsilon_{3z^2-r^2}-\varepsilon_{x^2-y^2}=150$ meV.

The inter-orbital hybridization also gives rise to two effects. 
The first one, which can be understood
intuitively, is an effective broadening of the more narrow band (see  Fig.~\ref{MOD_nonintDOS} for the
noninteracting case). This leads to a more metallic behavior of the
$3z^2-r^2$--electrons. The
second effect is more intrinsic in the sense that the hybridization
makes the bands more similar, obviously pushing the system towards
situation (i). It is important to notice that both effects work
against the effective crystal-field splitting.
As mentioned above, the fingerprints of the hybridization can clearly
be seen also in Fig.~\ref{MOD_spectra2D}: Although the narrow
$3z^2-r^2$--like--band is shifted above the Fermi energy, at the metal--insulator 
transition some residual $3z^2-r^2$ weight remains in the
region of the lower $x^2-y^2$--like--Hubbard band.

\subsubsection*{Double occupancy}
As in the case of the half-filled Hubbard model, complementary
information about the metal-insulator transition can be extracted from
analysis of the double occupancy $d$, which is the derivative of
the  free energy with respect to the interaction
parameter $U$ and is a good indicator for 
the Mott metal--insulator transition\cite{footnote02}. However, such an analysis for a
two-band model is more complicated since it involves not only the
orbital-diagonal double occupations ($d_{mm}=\langle n_{m\uparrow}
n_{m\downarrow}\rangle$) but also the orbital-offdiagonal parts with
parallel ($d^{\uparrow\uparrow}_{pa}=\langle n_{p\uparrow}
n_{a\uparrow}\rangle$) and antiparallel spin orientations
($d^{\uparrow\downarrow}_{pa}=\langle n_{p\uparrow}
n_{a\downarrow}\rangle$ ).

\begin{figure}
\begin{center}
  \includegraphics[width=7.5cm]{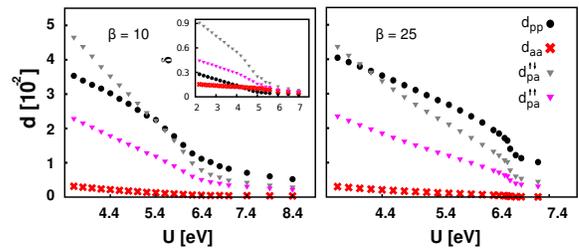}
  \caption{(Color online) Double occupancy vs.\ Coulomb interaction $U$ for two values of $\beta$ (left: $10$eV$^{-1}$; right: $25$eV$^{-1}$).
    Shown are the double occupation of the $x^2-y^2$ orbital ($d_{pp}$), of the  $3z^2-r^2$ orbital ($d_{aa}$) as well as the
    double occupation of both orbitals with parallel ($d^{\uparrow\uparrow}_{pa}$)
    and antiparallel spin ($d^{\uparrow\downarrow}_{pa}$). Inset: double occupancies normalized by the respective density ($\delta_{mn}$ see text).}  
  \label{MOD_NdPlot}
\end{center}
\end{figure}

The evolution of these four quantities is shown in
Fig.~\ref{MOD_NdPlot} for two different temperatures. In both cases we
observe the same trend for all double occupations which are decreasing
with increasing interaction. Starting the analysis with the
$x^2-y^2$-orbital, which undergoes a Mott-Hubbard transition when getting half filled, we
observe that its double occupation decreases around
$U=6.4\,$eV, i.e., near the critical value of the interaction that we can
estimate form the spectral function in Fig. \ref{MOD_spectra2D}. At
kT$=1/10$eV (Fig. \ref{MOD_NdPlot} left panel), however, this
behavior is thermally smeared out. The Mott-Hubbard transition occurs in form of a
very smooth crossover. A better estimate can be made by going down
in temperature to kT$=1/25$eV (Fig. \ref{MOD_NdPlot} right panel),
where we can see a very steep drop of $d_{pp}$ marking the transition
point at $U\approx 7.0\,$eV where we also found $Z\rightarrow 0$ for
the $x^2-y^2$ band.

At a first glance the behavior of the other three double occupancies
$d_{aa}$, $d^{\uparrow\downarrow}_{pa}$, and
$d^{\uparrow\uparrow}_{pa}$ appear to be more difficult to interpret:
One could have expected that $d^{\uparrow\uparrow}_{pa}$ and
$d^{\uparrow\downarrow}_{pa}$ should be the largest ones because they are
energetically less expensive ($V$,$V-J$) in comparison with the $d_{mm}$
($U$). What we observe, however, is a crossing of
$d^{\uparrow\uparrow}_{pa}$ with $d_{pp}$ at  $U\approx 5.4\,$eV
and $U\approx 4.2\,$eV for $\beta=10$ and $\beta=25$eV
respectively. Moreover, $d^{\uparrow\downarrow}_{pa}$ is lower than
$d_{pp}$  for all interaction values. This situation can be properly
understood considering also the depletion of the $3z^2-r^2$-band: The
low values of $d_{aa}$, $d^{\uparrow\downarrow}_{pa}$,  and
$d^{\uparrow\uparrow}_{pa}$ reflect the low $3z^2-r^2$-electron density.
In order to disentangle these effects we have plotted
the double occupancies normalized by the respective density in the
inset of the left panel of Fig. \ref{MOD_NdPlot} ($\delta_{mm} =
d_{mm}/\langle n_m \rangle^2$,
$\delta^{\uparrow\downarrow}_{pa}=d^{\uparrow\downarrow}_{pa}/(\langle
n_{p\uparrow} \rangle\langle n_{a\downarrow} \rangle)$,  and
$\delta^{\uparrow\uparrow}_{pa}=d^{\uparrow\uparrow}_{pa}/(\langle
n_{p\uparrow} \rangle\langle n_{a\uparrow} \rangle)$. In this way the
hierarchy of the energetic consideration is restored and all renormalized double occupancies behave in a similar way.
On the other hand, the crossing observed in the unrenormalized data
can be also interpreted as a clear signal that we are far from the
situation (i) described above, that is, a simultaneous Mott-Hubbard
transition in both bands.

\section{From 2D to 3D}
\label{Sec:3d}
Having discussed in detail the results for the two--dimensional model for the LaAlO$_3$/LaNiO$_3$ superlattice, we now present results for a 3D model for bulk LaNiO$_3$, and compare with the 2D case. Specifically, we no
longer restrict the hopping of the axial $3z^2-r^2$--orbital to the $xy$--plane, but
allow for hopping along the $z$--axis by adding the term
\begin{equation*}
  -2\cdot t_z\cdot \cos(k_z)\cdot
  \begin{pmatrix}
    0    & 0 \\
    0    & 1
  \end{pmatrix}
\end{equation*}
to the kinetic part of our Hamiltonian~\eqref{MOD_disp2D}.
We choose $t_z=0.6 {\rm eV}$, i.e., $4/3$ times the $0.45$ eV in equation~\eqref{MOD_disp2D}, as it should be in cubic symmetry according to the Slater-Koster tables\cite{slaterkoster} and Ref.~\onlinecite{Andersen78}. Our 3D model is not completely cubic because from equation~\eqref{MOD_disp2D} we kept the tetragonal crystal--field splitting (last term in Eq.~\eqref{MOD_disp2D}) as well as the second--nearest neighbor hoppings in $xy$--plane, without including them in the $xz$- and $yz$--planes. Moreover, this 3D model is not to be seen as a
true LDA+DMFT analysis of pseudo--cubic LaNiO$_3$, albeit the link to this compound is
a motivation for the analysis of the modified model.

\begin{figure}
  \begin{center}
    \includegraphics[width=7.5cm]{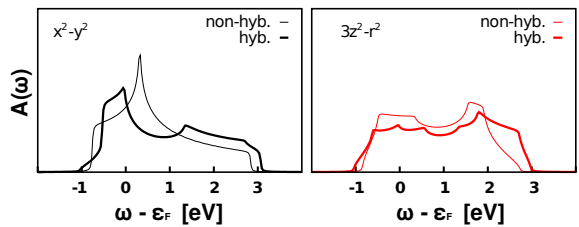}
    \caption{(Color online) Same noninteracting projected DOS as Fig. \ref{MOD_nonintDOS3D}, but for the 3D case. 
      The $3z^2-r^2$ band is considerably broader than in the layered 2D
      case due to the additional hopping in the z--direction.} 
    \label{MOD_nonintDOS3D}
  \end{center}
\end{figure}

The noninteracting density of states for this 3D model is
shown in Fig. \ref{MOD_nonintDOS3D}. The main difference from the 2D model
in Fig. \ref{MOD_nonintDOS} is obviously, a much broader
$3z^2-r^2$-band caused by the additional hopping in the 3rd direction. As a result of this broadening the two different
bands are much more similar than in the 2D case. In the truly cubic case 
the $x^2-y^2$ and $3z^2-r^2$ spectral densities would have been identical.

\begin{figure}
\begin{center}
	\includegraphics[width=7.5cm]{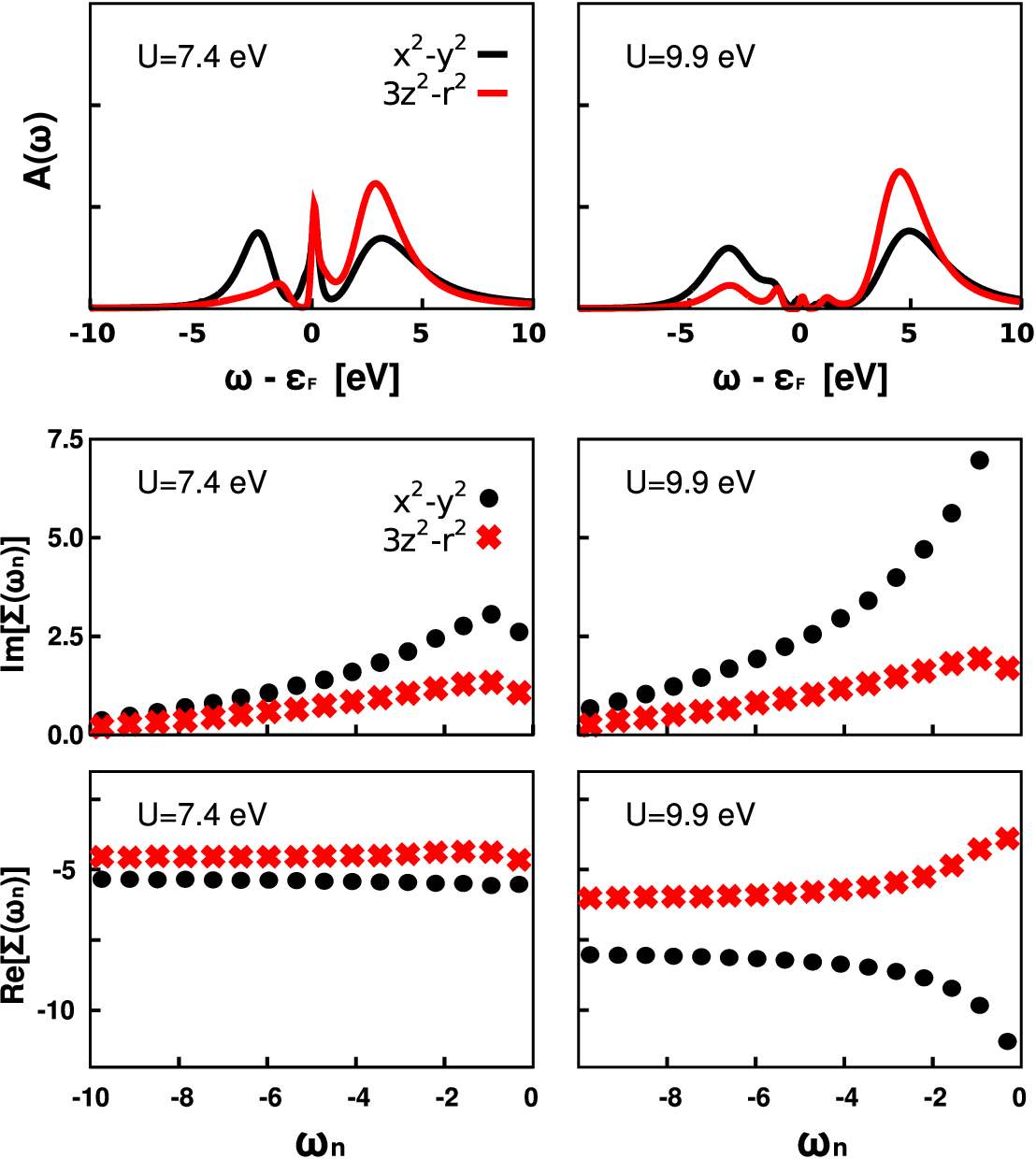}
  \caption{(Color online) Evolution of the spectrum and the
    LDA+DMFT self energy (vs.\ Matsubara frequencies) with increasing
    Coulomb interaction $V=U-2J$ for the 3D case at $\beta=10$eV$^{-1}$. The black
    curves show the $x^2-y^2$ spectrum and the red/gray ones the $3z^2-r^2$
    spectrum.}
  \label{MOD_spectra3D}
\end{center}
\end{figure}

In Fig.~\ref{MOD_spectra3D} we show the spectral functions and self
energies of the interacting 3D system. On the left hand (right hand) side
the data for a metallic (insulating) solution are plotted. They refer
to $U=7.4$ and $U=9.9\,$eV, respectively. This indicates that the metal--insulator 
transition occurs for considerably 
larger values of the interaction than
in the 2D system, as might be expected. Another difference
to the 2D case is the enhanced spectral weight of the
$3z^2-r^2$-band below the Fermi energy. 
However, the metal--insulator transitions for the two bands are
qualitatively similar to those of the 2D case as is clearly demonstrated
by the behavior of the imaginary part of the selfenergy shown in the
bottom panels of Fig. \ref{MOD_spectra3D}: The evolution of the self
energies qualitatively resembles the one shown in Fig. \ref{MOD_self2D}
displaying the same ``metallic'' down--bending. The only (quantitative)
difference to be noticed w.r.t. the 2D case is a larger imaginary
part of the $3z^2-r^2$-selfenergy at low-frequencies.
This can be understood as a natural consequence of the closer
resemblance of the bands in the noninteracting density of states. In
terms of our precedent discussion we thus observe again the situation (ii) of section \ref{Sec:MIT}, i.e., a simultaneous albeit quantitatively different Mott transition. However, we are now  slightly
closer to the case (i) of a simultaneous double Mott transition
for both bands than for the layered model.
In other words, the Mott transition is qualitatively the same for the 3D as 
 the 2D case, but quantitatively the Mott transition is shifted
to much higher values of the interaction strength. Such large values can hardly be realized in experiment.

It must however be emphasized that for a truly cubic 3D model, the two $e_g$ orbitals remain identical (for all strengths of the interaction) so that the Mott transition will occur simultaneously for both bands and for a value of U which is about $\sqrt 2$ larger than for the single-band case\cite{koch99}. That a small symmetry breaking at the LDA level, as exhibited by our 3D $e_g^1$ model, can drastically change the nature of the Mott transition in the paramagnetic state has previously been found in DMFT calculations for 3D $t_{2g}^2$ systems\cite{keller04,paper2} and $t_{2g}^1$ systems\cite{pavarini05}, as well as for the $e_g^1$ system LaMnO$_3$\cite{yamasaki06}.

\section{Predominating spin fluctuations and effective superexchange model}
\label{Sec:spinfluct}
\subsection*{Predominating spin fluctuations}
\begin{figure}
\begin{center}
  \includegraphics[width=7.5cm]{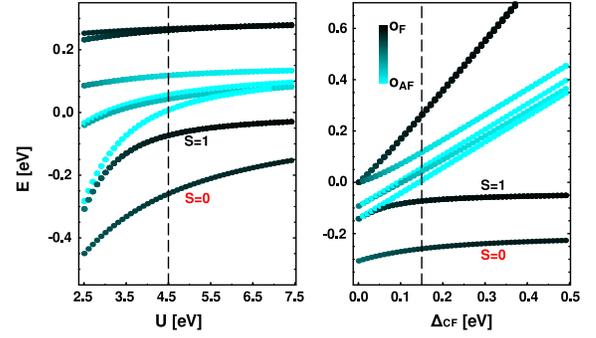}
  \caption{(Color online) Left hand side: Energy level diagram of the two site model as a
    function of the Coulomb repulsion $U$ for $\Delta_{CF}=0.15$eV (dashed line
    in the plot on the right-hand side). Right hand side (reproduced from Ref.~\onlinecite{Hansmann09}): Same diagram as a
    function of $\Delta_{CF}$ for $U=4.5$eV (dashed line in the plot on the
    left-hand side). The color of the points indicates the character
    of orbital ordering (o$_F$: orbital ferro; o$_{AF}$: orbital antiferro)}
  \label{MOD_levelplot}
\end{center}
\end{figure}

The model calculations discussed in the previous sections were performed
for the paramagnetic phase. However, one expects the system to display
magnetic and/or orbital order at sufficiently low temperatures. The treatment
of such complex ordered phases is particularly difficult in LDA+DMFT. An
insight into this physics can be obtained however by diagonalizing a ``two-site version'' of Hamiltonian
\eqref{eq:Heff} (with open boundary conditions, i.e., a diatomic molecule directed along the $x$-axis). The analysis of the
corresponding energy levels for this two-electron system allows to infer the relevant fluctuations
which dominate the low temperature physics. On the
left-hand side of Fig.~\ref{MOD_levelplot} we show the
calculated energy level diagrams as a function of U for the same set of parameters that we have
used for the LDA+DMFT 2D crystal calculation discussed in previous sections.

For the values of U considered, the mixing of the lowest energy
levels with the energy--costly double--occupied states is very small,
i.e., these states are already above the energy window which we plot
in Fig.~\ref{MOD_levelplot}. 
More
interestingly, our results show that the ground state of
the two site system is always a spin singlet with the two electrons residing
predominantly in the same orbital ($x^2-y^2$) on each site. 
The splitting of the singlet ground state and the excited triplet is large, and, in
the large U limit, it can be regarded as an exchange integral of $J_{\text ex.}\approx200$ meV.
This indicates an instability
of the system at low temperatures towards a spin antiferro- and orbital
ferro-ordered state. We have also studied the effect of varying the
initial level (i.e., crystal field) splitting for a reasonable value of $U=5$eV (right-hand side of
Fig.~\ref{MOD_levelplot}). The main effect of an enhancement of the crystal
field splitting is to strengthen the tendency towards ferrotype orbital ordering (of the
$x^2-y^2$ and the $3z^2-r^2$ states) eventually leading to purely one-band physics, with only
the $x^2-y^2$ orbital involved.

\subsection*{Effective superexchange model}
A first analysis of the magnetic and orbital ordered phases in the
La$_2$NiAlO$_6$ heterostructure was carried out in Ref.~\onlinecite{chaloupka08} considering the following
superexchange Hamiltonian, defined for a bond $ij\parallel\gamma$ in
the NiO$_2$ plane:

\begin{equation}
  \label{MOD_SuHam}
  H_{ij}^{(\gamma)}=(R_{\sigma,\pm,\pm}+R_{\sigma,\pm,\pm}^{CT})(\frac{1}{2}\pm \hat{\tau}_{i}^{(\gamma)})(\frac{1}{2}\pm \hat{\tau}_{j}^{(\gamma)})\hat{P}_{\sigma,ij}
\end{equation}

with an implied sum over $\sigma=0,1$ and all combinations of
$\pm,\pm$. $\hat{P}_{\sigma,ij}$ represents the projector to a singlet
and a triplet state of two Ni$^{3+}$ $S=\frac{1}{2}$ spins while
$\frac{1}{2}\pm \hat{\tau}_{j}^{(\gamma)}$ selects the planar orbital
in the plane perpendicular to the $\gamma$ axis and the directional
orbital along this axis.

\begin{figure}
  \begin{center}
    \includegraphics[width=7.5cm]{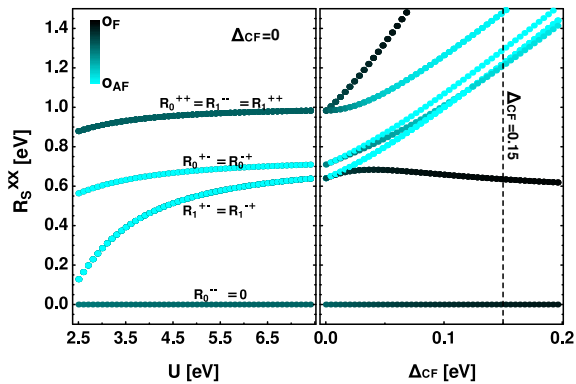}
    \caption{(Color online) Left hand side: Evaluation of the R-coefficients of
      Hamiltonian (\ref{MOD_SuHam}) as a function of the interaction parameter U
      calculated with the two site model in the $x^2-y^2$ and $3z^2-r^2$ basis
      for $\Delta_{CF}=0$eV . Right hand side: Evolution with increasing crystal
      field splitting starting from the largest U value of the left panel
      ($U=7.5$eV). The color of the points indicates the character
      of orbital ordering with the same scale as in Fig.\ \ref{MOD_levelplot}.}
    \label{MOD_Rplot}
  \end{center}
\end{figure}

Neglecting the charge transfer part of Hamiltonian (\ref{MOD_SuHam})
allows for a comparison with our two--site calculation, albeit only for
$\Delta_{CF}=0\,$eV since otherwise Hamiltonian (\ref{MOD_SuHam})
has another orbital symmetry than the full Hamiltonian.
In this way we can perform the first numerical
estimate of the $R_{\sigma,\pm,\pm}$ coefficients of the superexchange
Hamiltonian (\ref{MOD_SuHam}) as a function of the parameter of the
microscopic Hamiltonian. More specifically, the $R_{\sigma,\pm,\pm}$ coefficients of Fig.~\ref{MOD_Rplot}
are calculated form the level--splitting seen in Fig.~\ref{MOD_levelplot} which, in turn,
depends on the values of the parameters $U$, $J$, and $t$ of
Hamiltonian \eqref{eq:Heff}. The result of this comparison are
shown in the left panel of Fig.~\ref{MOD_Rplot}, where for simplicity we assumed the two orbitals to be degenerate.

It is important to note here, that the orbital projectors in
Eq.~(\ref{MOD_SuHam}) do not refer to the $x^2-y^2$ and $3z^2-r^2$ basis we
have used so far since the $\gamma$ axis is lying in the NiO$_2$ plane. This
is also reflected in the mixed color of the $R_{--}^0$ coefficient which
corresponds to the singlet ground state.

Finally, in the right panel we show the effect of turning on the crystal field
splitting $\Delta_{CF}$, disregarding some (here still small)  terms which
admix the orbitals in  Hamiltonian (\ref{MOD_SuHam}). 
The changes in the color clearly indicates a rapid
demixing of the eigenstates, since the $x^2-y^2$ and $3z^2-r^2$ orbitals are
eigenfunctions of the crystal field operator. This limits the applicability of
Hamiltonian (\ref{MOD_SuHam}) to the cases of small crystal field splittings.

As for the magnetic fluctuations, the Kugel--Khomskii--type Hamiltonian
(\ref{MOD_SuHam}) clearly favors an antiferromagnetic spin--alignment
along each pair. Depending on the crystal--field splitting, the
preferred orbital orientation is along the bond or --for larger
crystal field-- a ferro--orbital occupation of the $x^2-y^2$ orbital.

\section{Conclusion}
\label{Sec:conclusion}
We have analyzed in detail the Mott transition in the high-temperature, paramagnetic state of LaAlO$_3$/LaNiO$_3$ 
heterostructures, including the dimensional crossover
from the two-dimensional layered heterostructure to the three dimensional bulk 
system. The LDA bandstructure was NMTO-downfolded to the Ni $e_g$ orbitals
and, in order to concentrate on the essentials of the Mott physics, we 
 included merely the hoppings between first and second-nearest Ni neighbors. This realistic model also allowed us to contribute to the intensive
discussion on the orbital-selective Mott transition 
\cite{heldarita,moeller95,bulla99,rozenberg99,koga05,ferrero05,medici05,poteryaev07,anisimov02,bluemer09,medici09}, a discussion which with a few exceptions \cite{poteryaev07} has hitherto neglected the effect of interband-hybridization. This hybridization turned out
to be most important for the itinerancy of the $3z^2-r^2$ band and for
a simultaneous, albeit different Mott transition in both orbitals.

For realistic $U$ values we find that LaNiO$_3$/LaAlO$_3$ is a metal with an $(x^2-y^2)$--like quasiparticle peak at the Fermi level and a $(3z^2-y^2)$--like peak slightly above. At larger values of the Coulomb interaction,
the $3 z^2-r^2$ spectral density is shifted above the Fermi  as in
a {\it metal} to {\it band--insulator} transition for this band, while the  $x^2\!-\!y^2$--like band undergoes
a genuine Mott transition with divergent effective mass ($1/Z$).
Due to the hybridization between both orbitals, there is however still
spectral weight in the  $3 z^2-r^2$ channel 
at the position of the predominately  $x^2\!-\!y^2$ lower Hubbard band.
%This way the system self dopes the $x^2\!-\!y^2$ orbital off half-filling,
%an interesting aspect concerning the question of superconductivity.

If LaAlO$_3$/LaNiO$_3$ is grown on a substrate, stress can further enhance
 the initial (LDA) orbital splitting so that
one could obtain a situation which is more  single-band-like.
This is expected to reduce the critical Coulomb
interaction strength for the Mott transition.

Studying the two-site version of the model, which also includes 
second order processes in the nearest-neighbor hopping, we made
contact with an effective Kugel-Khomskii-Hamiltonian, indicating
predominant antiferromagnetic spin fluctuations, an essential ingredient 
for a prospective superconductor. Antiferromagnetic long-range
order is hence also possible and might enhance the tendencies towards an insulating ground state.

%\begin{acknowledgement}
We thank  G. Khaliullin for valuable discussions and
for initiating our interest in this system.
This work was supported in part by the Austrian Science Fund
(FWF) through SFB F41 (ViCoM), WK004 and Research unit FOR 1346 as well as
the EU-Indian network MONAMI and the National Science Foundation under Grant No. PHY05-51164 (KITP).
 Calculations have been performed
on the Vienna Scientific Cluster.
%\end{acknowledgment}


\begin{references}
\bibitem{footnote01}
Note, in the case of the LaAlO$_3$/LaNiO$_4$ heterostructures, the $3z^2-r^2$
orbital points towards the insulating AlO layers. Since these are
insulating the hopping in $z$--direction is strongly reduced and can
approximately be neglected.
\end{references}
\end{document}